\input{aipcheck}

\documentclass[
    ,final            
,sort&compress]{aipproc}

\layoutstyle{8x11single}
\usepackage{amsfonts}

\begin{document}

\title{Wave Transport in disordered waveguides: closed channel contributions
and the coherent and diffuse fields}



\author{M. Y\'epez}{
address={Departamento de F\'isica, Universidad Aut\'onoma
Metropolitana-Iztapalapa, Apartado Postal 55-534, 09340 M\'exico Distrito
Federal, Mexico}
}

\begin{abstract}
We study the wave transport through a disordered system inside a waveguide. The
expectation value of the complex reflection and transmission coefficients (the
coherent fields) as well as the transmittance and reflectance are obtained
numerically. The numerical results show that the averages of the coherent fields
are only relevant for direct processes, while the transmittance and reflectance
are mainly dominated by the diffuse intensities, which come from the statistical
fluctuations of the fields.
\end{abstract}

\maketitle


\section{Introduction}
\label{Sec:Intro}

The wave transport through disordered systems has been of great interest in
many fields of physics, where the wave interference phenomena plays a
transcendental
role~\cite{Landauer:PhilMag,Ishimaru:1978,Altshuler:1991,Sheng:1995,Imry:1997,
Datta:1997,RevModPhys.69.731,Mello:2010}. The interference phenomena that occur
when a wave propagates through a disordered medium containing a random
distribution of scatterers is so complex that any change in the microscopic
realization of the disorder modifies completely the interference pattern of a
macroscopic observable~\cite{Stone:1988}; therefore, only a statistical
description makes sense. The complexity derives from the fluctuations in the
refractive index as in disordered dielectric medium, the randomness of the
scattering potentials, as in the case of disordered conductors with impurities
or more in general, in disordered waveguides; the complexity no only derives
from the randomness of the system, it is also consequence of the multiple
scattering processes. In the present work we focus the study in the domain of
disordered waveguides and in the quantum mechanics context (electron or scalar
waves), nevertheless, the method could be applied to classical waves:
electromagnetic or elastic waves.

Some previous studies in {\em quasi-one dimensional} (Q1D) disordered systems
have found remarkable regularities for the statistical properties of wave
transport in the sense that the statistic of macroscopic observables involves a
rather small physical parameters, {\em the mean free paths} (MFPs). Those models
are in good agreement with the scaling
approaches~\cite{AnnaPhys.181.290,PhysRevB.37.5860,PhysRevLett.89.246403,
PhysRevB.46.15963, PhysRevE.75.031113,PhysRevB.83.245328}, including the
celebrated Dorokhov-Mello-Pereyra-Kumar
(DMPK)~\cite{AnnaPhys.181.290,JETPLett.36.318} and the non linear
sigma-model~\cite{JETPLett.58.444,IntJMP.8.3795,PhysRevB.53.1490} approaches. 

Most of previous works were mainly focus on the study of the statistics of the
transport coefficients, i.e., the transmittances, reflectances and the
dimensionless conductance (transmission intensity), while the statistical
properties of the complex coefficients or {\em coherent fields}, were not
studied in detail; moreover, in those previous approaches, the closed channels
or evanescent waves are not included in the description. For instance, the DMPK
approach describes successfully the statistical properties of the conductance,
where the only relevant physical parameter is the transport or elastic mean free
path $\ell$; however, the DMPK model is not suitable to describe the statistics
of the complex transmission and reflection coefficients. The models developed in
Refs.~\cite{PhysRevB.46.15963,PhysRevE.75.031113} give a more general
description than the DMPK approach. In those models, the macroscopic statistics
only depends on the channel-channel mean free paths $\ell_{aa_{0}}$ and the
scattering mean free path $\ell_{a_{0}}$: $a$ and $a_{0}$ denote, respectively,
the modes or channels of the incoming and outgoing waves. In principle, the
approaches given Refs.~\cite{PhysRevB.46.15963,PhysRevE.75.031113} are
appropriate to describe the statistics of the coherent fields; however, those
models do not consider the closed contributions, which, as it is demonstrated in
Ref.~\cite{YepezSaenz:2013}, play a transcendental role in the statistical
properties of the complex transmission and reflection coefficients.

In the present work we analyze numerically the influence of the closed channels
in the macroscopic statistics of disordered waveguides. We present numerical
results for the expectation values of the complex reflection and transmission
coefficients, as well as of the corresponding intensities. In addition, we also
present numerical simulations for coherent and diffuse intensities, which have
not been studied in detail in previous theoretical and numerical studies. For
that purpose we will use the {\em extended} or {\em generalized scattering
matrix} (GSM) technique~\cite{Mello:2010,Tores_Saenz_JPSJ.73.2182,Mittra:1971,
mittra1988techniques, Weisshaar_JAP_70_355_1991}.

\section{Statistical Scattering Properties of disordered waveguides} \label{Sec:Theory}

\subsection{Generalized and reduced scattering matrices}

\begin{figure}
\includegraphics[scale=0.37]{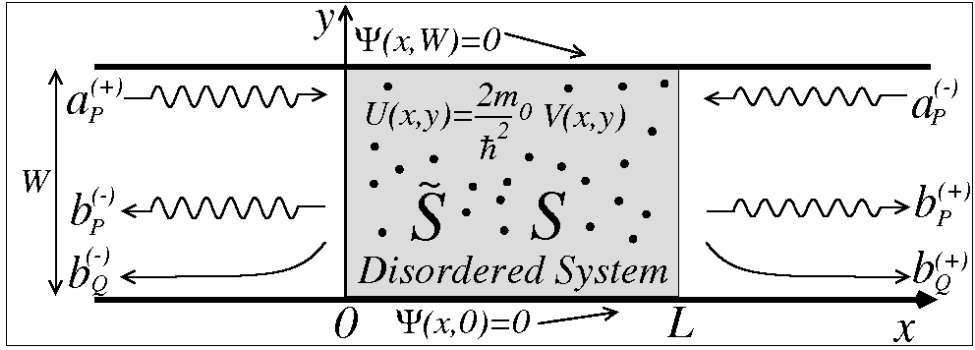}
\caption{Scattering problem in a waveguide. $a_{P}^{\left(+\right)}$ and
$a_{P}^{\left(-\right)}$ denote $N$ dimensional vectors, being their elements
all possible incoming open channel amplitudes. Analogously
$b_{P}^{\left(-\right)}$ and $b_{P}^{\left(+\right)}$ are vectors with all
possible outgoing open channel amplitudes and $b_{Q}^{\left(-\right)}$,
$b_{Q}^{\left(+\right)}$, are the corresponding outgoing closed channel vectors,
whose dimensionality $N^{\prime}$ is, in principle, infinite.}
\label{Theregimes} 
\end{figure}

Consider a two dimensional disordered system of uniform cross section and length
$L$ inside a waveguide with impenetrable walls, a constant width $W$, that is
clean at both sides of the disordered region (see Fig. \ref{Theregimes}). The
disordered system, which hereafter shall be called the {\em Building Block}
(BB), is represented by a random potential $U$, whose microscopic model is
introduced in the next section. Inside the waveguide, the solution of the wave
equation $\nabla^{2}\Psi+k^{2}\Psi=U\Psi$ ($k$ being the wavenumber in the
clean region) is written as a series of traveling and evanescent waves, which
are associated to {\em open and closed channels}, respectively. The waveguide
supports precisely $N$ open channels or traveling modes when $N<kW/\pi<N+1$,
while the number of closed channels or evanescent modes $N^{\prime}$ is, in
principle, infinite.

In Fig.~\ref{Theregimes} it is shown the most general situation of the
scattering problem, where incoming-waves of open channels
\begin{small}$a_{P}^{\left( + \right) }$\end{small} and \begin{small}$a_{P}^{
\left(- \right) }$\end{small} [incoming-waves in closed channels are not
allowed, so \begin{small}$a_{Q}^{\left(+\right)}=a_{Q}^{\left(
-\right)}=0$\end{small}], are scattered by the disordered system, giving rise to
outgoing-waves, both in open channels (traveling modes) \begin{small}$b_{P}^{
\left(+\right) }$\end{small}, \begin{small}$b_{P}^{ \left(-\right)}$\end{small}
as in closed channels (evanescent modes) \begin{small}$b_{Q}^{ \left(+ \right)
}$\end{small}, \begin{small}$b_{Q}^{ \left(-\right)}$\end{small}; the symbols
$+$ and $-$ denote, respectively, waves traveling to the right and to the left,
while $P$ and $Q$ represent open and closed channel components, respectively.
The scattering problem is formally described by GSM
$\widetilde{S}$~\cite{Mello:2010}, which relates open and closed channel 
outgoing-wave amplitudes to the open channels incoming-wave amplitudes, i.e., 
\begin{equation}
\left(\begin{array}{c}
b_{P}^{\left(-\right)} \\
b_{Q}^{\left(-\right)} \\
b_{P}^{\left(+\right)} \\
b_{Q}^{\left(+\right)}
\end{array}\right)
=\widetilde{S} \left(\begin{array}{c}
a_{P}^{\left(+\right)} \;\;\;\;\;\;\\
a_{Q}^{\left(+\right)}=0 \\
a_{P}^{\left(-\right)} \;\;\;\;\;\;
\\
a_{Q}^{\left( -\right)}=0
\end{array}\right), \quad\mbox{with}\quad
\widetilde{S}=\left(\begin{array}{cc}
\widetilde{r} & \widetilde{t}'\\
\widetilde{t} & \widetilde{r}'
\end{array}\right).
\label{MyPHDTEc2_92a_b}
\end{equation}

Even though the $\widetilde{S}$ matrix describes completely the scattering problem, it is important to notice that the closed channel amplitudes \begin{small}$b_{Q} ^{\left(-\right)}$\end{small}, \begin{small}$b_{Q}^{\left(+\right)}$\end{small} decrease exponentially as we move away from the disordered system, which is shown schematically in Fig. \ref{Theregimes}; moreover, those amplitudes do not contribute to the flux density current. For this reason, the scattering problem is usually described in terms of the well known {\em open channels} or {\em reduced scattering matrix} $S$, that relates open channel outgoing- and incoming-wave amplitudes in the asymptotic region, i.e.,

\begin{equation}
\left( \begin{array}{c}
b_{P}^{\left(-\right)}
\\
b_{P}^{\left(+\right)}
\end{array}
\right)=S
\left(
\begin{array}{c}
a_{P}^{\left(+\right)}
\\
a_{P}^{\left(-\right)}
\end{array}
\right), \quad\mbox{with}\quad
S=
\left(
\begin{array}{cc}
r & t^{\prime}
\\
t & r^{\prime}
\end{array}
\right).
\label{ScatMat_Relat}
\end{equation}

\subsection{Generalized scattering matrix technique}\label{Exte_S_Matrix}

In order to study the statistical scattering properties of the BB, we consider
an initial condition in which an incoming-wave in the open channel $a_{0}$
travels from left to right, so that the outgoing-waves (backward and forward)
are generated in any possible open channel $a=1,\cdots,N$: from now on $a_{0}$
and $a$ will only denote, respectively, the incoming and outgoing open channels.
Under this initial condition, we are interested in the statistics of observables
related to the complex reflection $r_{aa_{0}}$ ($\in r$) and transmission
$t_{aa_{0}}$ ($\in t$) coefficients as well as the corresponding reflectance
$R_{aa_{0}}=\vert r_{aa_{0}} \vert^{2}$ and transmittance $T_{aa_{0}}=\vert
t_{aa_{0}} \vert^{2}$, and the dimensionless conductance
$g=\sum_{a,a_{0}}^{N}T_{aa_{0}}$, which has been widely studied.

The statistical scattering properties of the BB, are obtained numerically from
an ensemble, where each member represents any possible microscopic realization
of the disorder. For instance, the expectation values $\left\langle
r_{aa_{0}}\right\rangle $, $\left\langle t_{aa_{0}}\right\rangle $,
$\left\langle R_{aa_{0}}\right\rangle $ and $\left\langle
T_{aa_{0}}\right\rangle $ are calculated as the average over the ensemble, where
the {\em generalized scattering matrix technique} is used. To illustrate how
this method is implemented, it is important to notice that each element of the
reduced matrix $S$, Eq.~\eqref{ScatMat_Relat}, is extracted from the generalized
matrix $\widetilde{S}$, Eq.~\eqref{MyPHDTEc2_92a_b}, which in turn is calculated
by combining the generalized scattering matrices of the individual scatters of
the system: see details of the combination law in Ref.~\cite{Mello:2010}. The
combination law captures any possible open or closed channel transition, which
are result of the multiple scattering processes inside the disordered system.
Once the reduced matrix $S$ is known, the complex reflection $r_{aa_{0}}$ and
transmission $t_{aa_{0}}$ coefficients, as well as the corresponding reflectance
$R_{aa_{0}}=\vert r_{aa_{0}} \vert^{2}$ and transmittance $T_{aa_{0}}=\vert
t_{aa_{0}} \vert^{2}$, are easily calculated. This procedure is repeated
numerically for each microscopic realization of the ensemble, what allows us to
generate an ensemble of matrices $S$ and consequently to obtain, numerically,
the ensemble averages $\langle r_{aa_{0}} \rangle$, $\langle t_{aa_{0}}
\rangle$, $\langle R_{aa_{0}} \rangle$ and $\langle T_{aa_{0}} \rangle$.

The implementation of the GSM method mentioned above, exhibits that the closed
channel or evanescent waves contribute implicitly to the $S$ matrix given in
Eq.~\eqref{ScatMat_Relat}. For this reason the $S$ matrix of any realization of
the microscopic disorder - and consequently the statistical properties
associated to $S$ - depend formally on the closed channels, which in general are
neglected in the theoretical studies of quasi one dimensional (Q1D) disordered
systems. 

The GSM method also guaranties the flux conservation property $S^{\dag}S=I$, which imposes the condition 
\begin{equation}
T_{a_{0}}+R_{a_{0}} =1,
\label{FluxConProp}
\end{equation}
being 
\begin{equation}
T_{a_{0}}=\sum_{a=1}^{N} T_{aa_{0}},\;\;\;\;\, R_{a_{0}}=\sum_{a=1}^{N}
R_{aa_{0}},
\end{equation}
the total transmittance $T_{a_{0}}$ and reflectance $R_{a_{0}}$, respectively;
the additions on $a$ is over any possible outgoing open channel. Since any
microscopic realization of the disorder satisfies the flux conservation
property, Eq.~\eqref{FluxConProp}, then the statistical properties will also be 
consistent with this general condition.

\subsection{Coherent and diffuse fields}

For a given realization of the microscopic disorder the complex coefficients of the transmitted and reflected waves can be written as the sum of the average $\left\langle t_{aa_{0}}\right\rangle$, $\left\langle r_{aa_{0}}\right\rangle$ (coherent) and residual $\Delta t_{aa_{0}}$, $\Delta r_{aa_{0}}$ (diffuse) fields, i.e.,
\begin{eqnarray}
t_{aa_{0}}=\left\langle t_{aa_{0}}\right\rangle + \Delta t_{aa_{0}},
& & \langle \Delta t_{aa_{0}} \rangle \equiv 0,
\\
r_{aa_{0}}=\left\langle r_{aa_{0}}\right\rangle + \Delta r_{aa_{0}},
& & \langle \Delta r_{aa_{0}} \rangle \equiv 0.
\end{eqnarray}
$\Delta t_{aa_{0}}$, $\Delta r_{aa_{0}}$ give the statistical fluctuations
around the corresponding coherent fields $\langle t_{aa_{0}}\rangle$, 
$\langle r_{aa_{0}}\rangle$. In a similar way, the transmittance and reflectance
of a given realization are written in the following way:
\begin{eqnarray}
T_{aa_{0}} = \vert t_{aa_{0}}\vert^{2} =
\left\langle T_{aa_{0}}\right\rangle + \Delta T_{aa_{0}},
& & \langle\Delta T_{aa_{0}}\rangle\equiv 0, \\
R_{aa_{0}} = \vert r_{aa_{0}}\vert^{2} = 
\left\langle R_{aa_{0}}\right\rangle + \Delta R_{aa_{0}},
& & \langle\Delta R_{aa_{0}}\rangle\equiv 0,
\end{eqnarray}
where 
\begin{eqnarray}
\langle T_{aa_{0}} \rangle &=& 
\vert\langle t_{aa_{0}} \rangle\vert^{2} + \langle \vert \Delta t_{aa_{0}}\vert
^{2}\rangle,
\label{Taa0Coehe_Diffuse_fiels}
\\
\langle R_{aa_{0}} \rangle &=& 
\vert\langle r_{aa_{0}} \rangle\vert^{2} + \langle \vert \Delta r_{aa_{0}}\vert
^{2}\rangle  ,
\label{Raa0Coehe_Diffuse_fiels}
\end{eqnarray}
denote, respectively, the expectation values of the transmittance and
reflectance coefficients, while
\begin{eqnarray}
\Delta T_{aa_{0}} &=& \vert \Delta t_{aa_{0}} \vert ^{2} -\left\langle \vert \Delta t_{aa_{0}}\vert ^{2}\right\rangle + 2\mathrm{Re}\left( \left\langle t_{aa_{0}}\right\rangle \Delta t_{aa_{0}}^{\ast} \right),
\\
\Delta R_{aa_{0}} &=&  \vert \Delta r_{aa_{0}} \vert ^{2} -\left\langle \vert \Delta r_{aa_{0}}\vert ^{2}\right\rangle +2\mathrm{Re}\left( \left\langle r_{aa_{0}}\right\rangle \Delta r_{aa_{0}}^{\ast} \right) ,
\end{eqnarray}
give the corresponding statistical fluctuations.
Equations~\eqref{Taa0Coehe_Diffuse_fiels} and~\eqref{Raa0Coehe_Diffuse_fiels}
show that the ensemble average of the transmittance 
$\left\langle T_{aa_{0}}\right\rangle$ and reflectance 
$\left\langle R_{aa_{0}}\right\rangle$ are constituted of two different
contributions: the coherent intensities 
$\vert\langle t_{aa_{0}}\rangle\vert^{2}$, 
$\vert\langle r_{aa_{0}}\rangle\vert^{2}$ and the diffuse intensities
$\langle\vert\Delta t_{aa_{0}}\vert ^{2}\rangle$, 
$\langle\vert\Delta r_{aa_{0}}\vert ^{2}\rangle$.

\section{\label{MicStatPotMod}Potential model for the Building Block}

\begin{figure}
\begin{tabular}{cc}
\includegraphics[height=4.3cm,width=8.0cm]{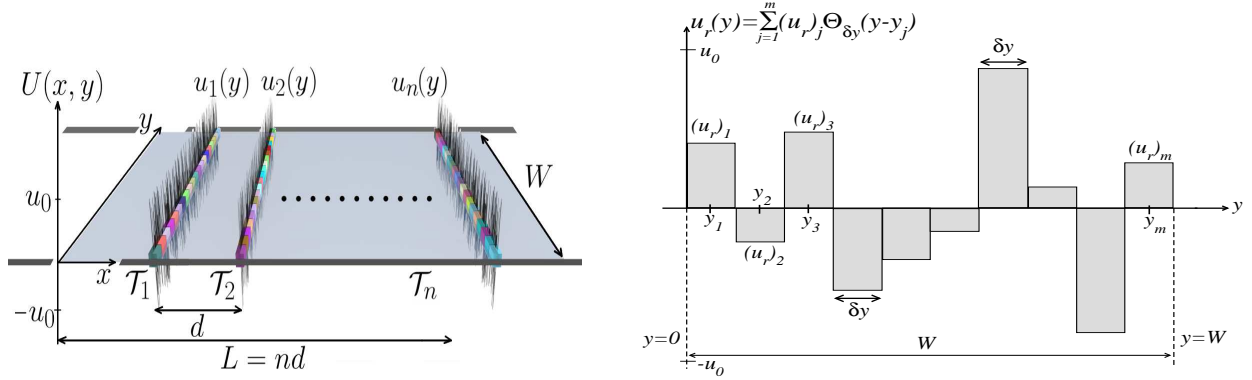} &
\includegraphics[height=5.0cm,width=8.0cm]{AIPConfProcYepez3.eps} 
\end{tabular}
\caption{Left: Schematic representation of the building block as sequence of
``thin'' $\delta$-potential slices. Right: Profile of the transverse dependence
$u_{r}\left( y\right)$, Eq. \eqref{MyPHDTEq6.13}, of the $r$th scattering
unit. The random values $\left(u_{r}\right)_{j}$ were generated by using the
uniform distribution of Eq.~\eqref{MicroDistri}.}
\label{deltasecuence}
\end{figure}

The Building Block is represented by a random potential $U\left(x,y\right)$ (in
units of $k^{2}$), that is constructed as a sequence of $n$ ($\gg 1$)
statistically independent and identically distributed scattering units, which
are separated from each other by a fixed distance $d$ in the wave propagation
direction $x$ [see Fig.~\ref{deltasecuence} (left)]. The scattering properties
of a scattering unit, are described by means of its extended scattering matrix
$\widetilde{s}_{r}$, which is proportional to the so called {\em transition
matrix} $\mathcal{T}_{r}$, that captures any possible channel-channel
transition~\cite{Messiah:1999,Newton:1982,Roman:1965}. The thickness of each
scattering unit is much smaller than the distance $d$, which in turn is much
smaller than the wavelength $\lambda$. The $r$th scattering unit
($r=1,2,\cdots ,n$) centered at $x_{r}=rd$, is specified by its potential
$U_{r}\left(x,y\right)$ (in units of $k^{2}$), that is approximated as a delta
potential slice in the longitudinal direction $x$, with random strength
$u_{r}\left(y\right)$ (in units of $k$) in the transverse direction $y$;
therefore, the BB is a system of length $L=nd$ given by the following potential
model:
\begin{equation}
U\left( x,y\right) =
\sum_{r=1}^{n}U_{r}\left(x,y\right)=\sum_{r=1}^{n}u_{r}
\left(y\right)\delta\left(x-x_{r}\right).
\label{Mic_Pot_Mod} 
\end{equation}

The microscopic model of a delta slice that is used in the present work is shown
in Fig.~\ref{deltasecuence} (right), where the transverse dependence of the
$r$th slice $u_{r}\left(y\right)$, Eq.~\eqref{Mic_Pot_Mod}, is generated in the
following way: i) The width of the waveguide $W$ is divided into $m\gg 1$
segments all of them with the same length $\delta y=W/m\ll\lambda$ (being
$\lambda$ the wavelength). ii) The $j$th ($j=1,2,\cdots m$) segment is
centered at $y_{j}=\left(j-1/2\right)\delta y$ and is defined by the interval 
$y\in\left[\left(j-1\right)\delta y,j\delta y\right]$. iii) Inside each
interval the function $u_{r}\left(y\right)$ takes a constant potential value
$\left( u_{r}\right)_{j}$, which is sampled from the uniform distribution
\begin{equation}
P(u_r)= \frac{1}{2u_{0}}, \quad \left( u_{r}\right)_{j}
\in\left[-u_{0},u_{0}\right].
\label{MicroDistri} 
\end{equation}

The procedure explained above generates the random profile for the $r$th
scattering unit, which is mathematically represented by the expression
\begin{equation} 
u_{r}\left( y\right) = 
\sum_{j=1}^{m}\left(u_{r} \right) _{j}\Theta _{\delta y}\left(y-y_{j}\right),
\label{MyPHDTEq6.13} 
\end{equation} 
where the $\Theta _{\delta y}\left(y-y_{j}\right)$ is the finite step 
function that takes the value 1 if 
$y\in\left[y_{j}-\delta y/2, y_{j}+\delta y/2\right]$ and zero if otherwise.


\section{Numerical results}

In this section we present numerical results for the expectation values
$\left\langle t_{aa_{0}}\right\rangle$, $\left\langle r_{aa_{0}}\right\rangle$,
$\left\langle T_{aa_{0}}\right\rangle$ and 
$\left\langle R_{aa_{0}}\right\rangle$ of the BB. The aim is to study their
evolution with $L$ and the role of the closed channels in the statistical
scattering properties of the Building Block. The influence of the closed
channels is analyzed numerically by considering four numerical simulations for
the expectation values of interest. These simulations are performed for a
waveguide that supports $N=2$ open channels (with $kW/\pi=2.5$), but each
simulation takes into account $N^{\prime}=0,1,2,3$ closed channels in the
calculations, respectively. The numerical expectation values 
$\left\langle t_{aa_{0}}\right\rangle$, $\left\langle r_{aa_{0}}\right\rangle$,
$\left\langle T_{aa_{0}}\right\rangle$, and 
$\left\langle R_{aa_{0}}\right\rangle$, are obtained as the average over an
ensemble of $10^{6}$ realizations of the microscopic disorder, where the
scattering matrix of each realization is generated by using the potential model
of the previous section and the GSM method introduced in the second section. As
in previous works, the numerical expectation values will be plotted as function
of the dimensionless length $L/\ell$, $\ell$ being the trasnport mean free path.

\begin{figure}
\begin{tabular}{cc}
\includegraphics[width=8.0cm]{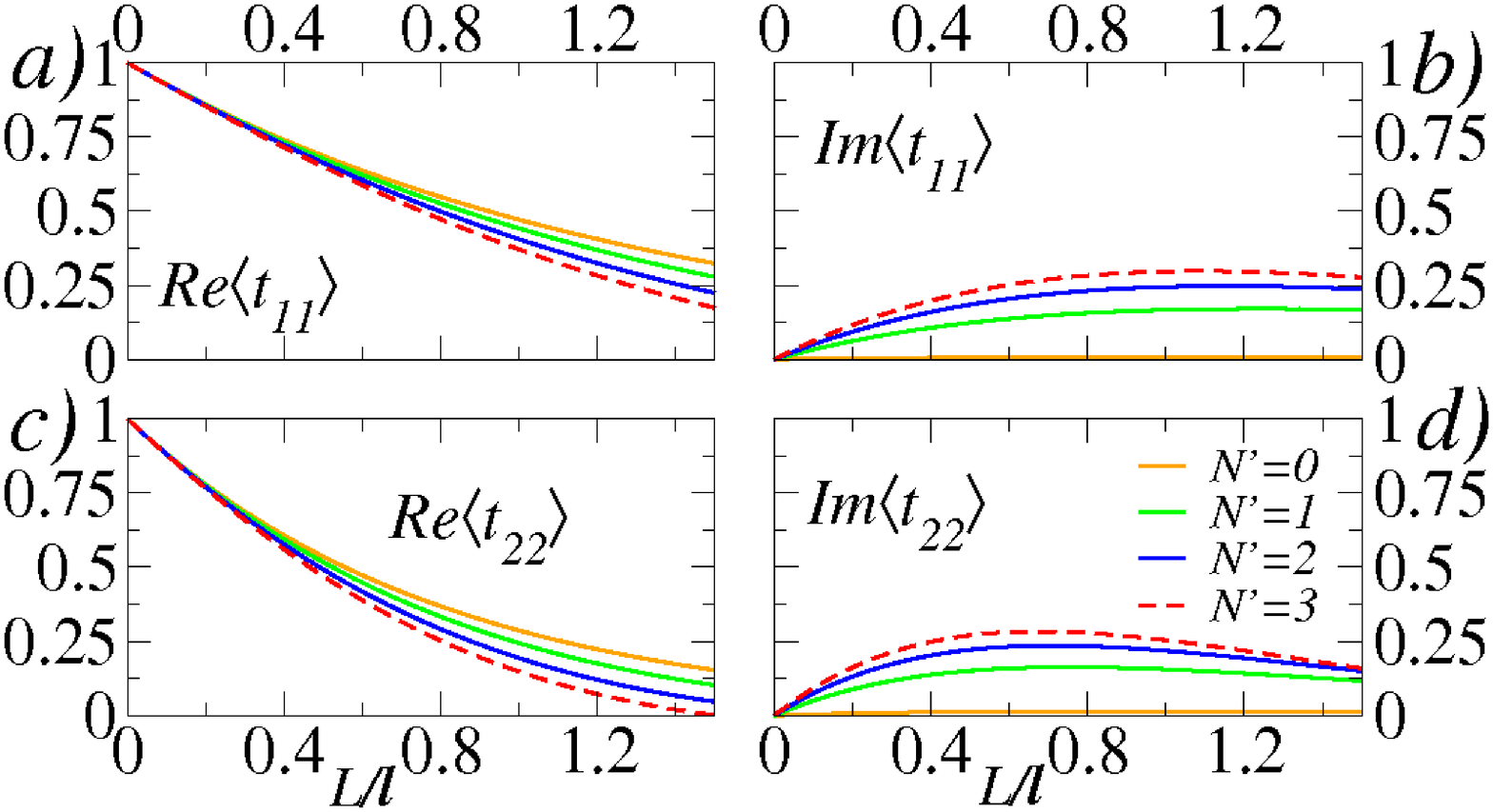} &
\includegraphics[width=8.0cm]{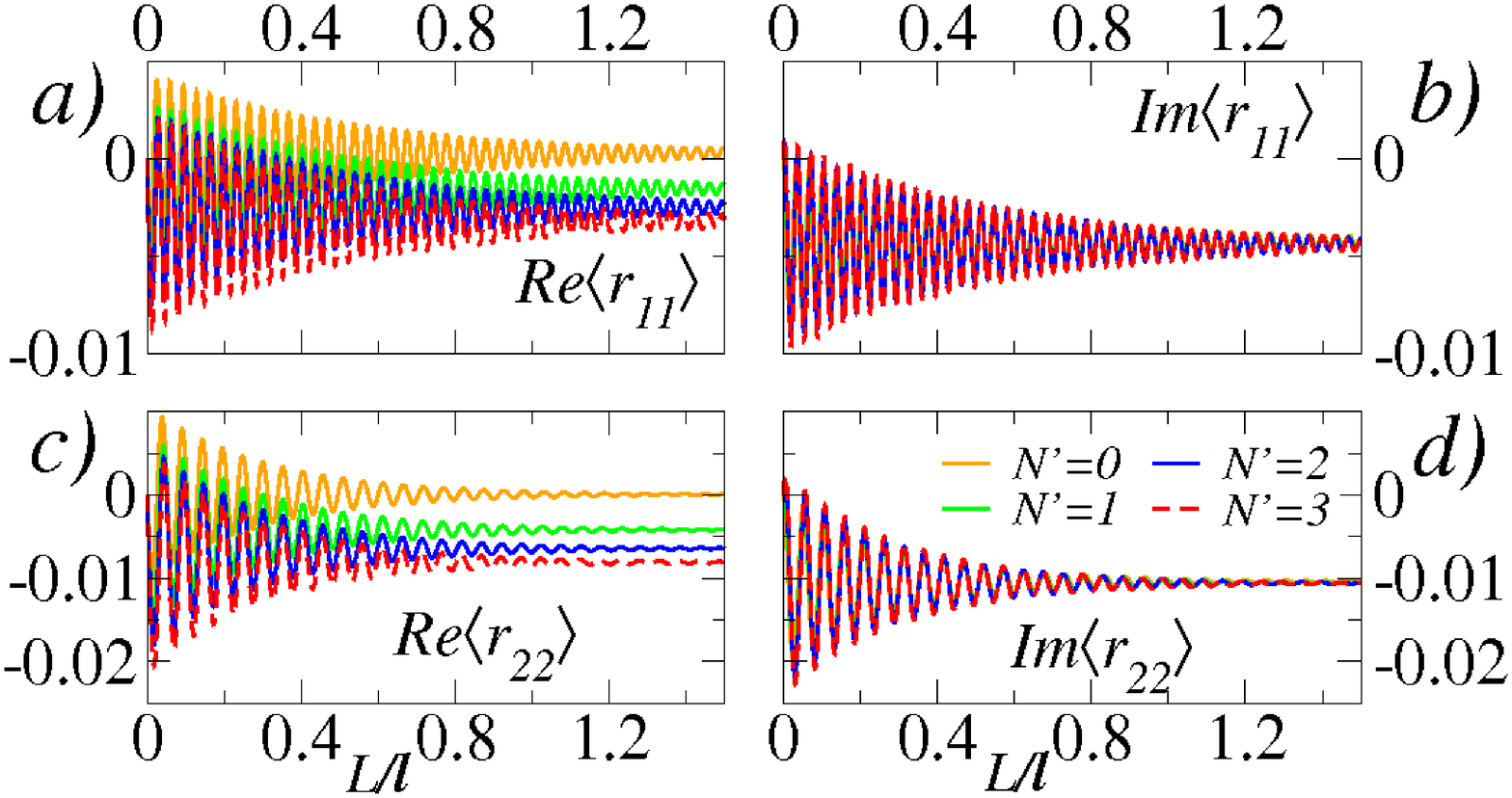}
\end{tabular}
\caption{Numerical results for $\left\langle t_{a_{0}a_{0}}\right\rangle$ (left)
and $\left\langle r_{a_{0}a_{0}}\right\rangle$ (right) vs $L/\ell$. Each
simulation considers $N=2$ propagating modes and several evanescent modes
($N^{\prime}=0,1,2,3$).}
\label{ScatteAmplit1}
\end{figure}

The numerical results of Fig.~\ref{ScatteAmplit1} (left) show that
$\mathrm{Re}\left\langle t_{a_{0}a_{0}}\right\rangle$ decreases as $L/\ell$
increases; this behavior becomes more notorious as the number of closed channels
($N^{\prime}=0,1,2,3$) considered in the calculations is increased. This figure
also shows that the influence of the closed channels in 
$\mathrm{Im}\left\langle t_{a_{0}a_{0}}\right\rangle$ is even more dramatic: if
closed channels are not included in the numerical simulations ($N^{\prime}=0$),
$\mathrm{Im}\left\langle t_{a_{0}a_{0}}\right\rangle $ is small 
($\sim 10^{-2}$), but not strictly zero; however, when the closed channels are
considered ($N^{\prime}=1,2,3$), 
$\mathrm{Im}\left\langle t_{a_{0}a_{0}}\right\rangle$ increases one order of
magnitude.

In the right panel of Fig.~\ref{ScatteAmplit1} we can appreciate a remarkable
oscillatory behavior for $\left\langle r_{a_{0}a_{0}}\right\rangle$, which is
rapidly attenuated as $L/\ell$ increases, regardless the number of closed
channels that are considered in the calculation. The phase and amplitude of
$\left\langle r_{a_{0}a_{0}}\right\rangle$ depend little on the closed channels
contributions. However, the most notorious dependence on the closed channel
contributions is exhibited by 
$\mathrm{Re}\left\langle r_{a_{0}a_{0}}\right\rangle$, whose oscillations are
given around a ``background'' that depends on the number of closed channels
considered in the calculations; in contrast, the four simulations of
$\mathrm{Im}\left\langle r_{a_{0}a_{0}}\right\rangle$ seem to oscillates around
the same ``background,'' no matter how many closed channels were used in the
calculations.


Most of previous theoretical models cannot describe the numerical evidence shown
in Fig.~\ref{ScatteAmplit1} for $\left\langle t_{a_{0}a_{0}}\right\rangle$ and
$\left\langle r_{a_{0}a_{0}}\right\rangle$, due to the absence of closed
channels. As an example, we consider Mello-Tomsovic (MT)
prediction~\cite{PhysRevB.46.15963} for the expectation values of the coherent
fields
\begin{eqnarray}
\left\langle t _{aa_{0}} \right\rangle^{\mathrm{(MT)}} &=& \delta_{aa_{0}} e^{-L/\ell_{a_{0}}},
\label{MyPHDEq7.89}
\\
\left\langle r _{aa_{0}} \right\rangle_{L}^{\mathrm{(MT)}} &=& 0,
\label{MyPHDEq5.33c} 
\end{eqnarray}
where $\ell_{a_{0}}$ is the scattering mean free path of the incoming open
channel $a_{0}$. We have verified that the MT prediction 
$\left\langle t_{a_{0}a_{0}}\right\rangle^{\mathrm{(MT)}}=e^{-L/\ell_{a_{0}}}$
is indistinguishable from the numerical result of 
$\mathrm{Re}\left\langle t_{a_{0}a_{0}}\right\rangle$ when the closed channels
are not considered in the calculation. However, once the closed channels are
included ($N^{\prime}=1,2,3$), 
$\mathrm{Re}\left\langle t_{a_{0}a_{0}}\right\rangle$ decreases faster than MT
predicts, Eq.~\eqref{MyPHDEq7.89}. In addition, the behavior 
$\mathrm{Im}\left\langle t _{a_{0}a_{0}}\right\rangle\neq 0$ is not predicted
by MT result. On the other hand, although the numerical results of
Fig.~\ref{ScatteAmplit1} (right) show that the amplitude of the oscillation of
$\left\langle r_{a_{0}a_{0}}\right\rangle$ is small ($\sim 10^{-2}$), this
expectation value is not zero, even when the closed channels are not included;
therefore, the MT prediction 
$\left\langle r_{a_{0}a_{0}}\right\rangle^{\mathrm{(MT)}}=0$ is not able to
describe the numerical results shown in right panel of Fig.~\ref{ScatteAmplit1}.

In previous studies~\cite{PhysRev.85.621,PhysRev.67.107}, it has been shown that
the coherent wave fields are characterized by an effective wave number
$k_{\mathrm{eff}}$, whose real part determines the phase of the coherent field,
while its imaginary part represents the losses due to scattering (often known as
waveguide extrinsic losses). As the wave propagates, the amplitude of the
coherent part decays exponentially with the length $L$ of the system. However,
in a recent study~\cite{YepezSaenz:2013} it has been theoretically
demonstrated that the exponential decay of the coherent fields 
$\left\langle t_{aa_{0}}\right\rangle$ and 
$\left\langle r_{aa_{0}}\right\rangle$ is strongly modified when the closed
channels are correctly taken into account: changes in the phase of the coherent
field (i.e. in the real part of the effective wave number) are solely related to
evanescent modes, while the scattering mean free path $\ell_{a_{0}}$ is
insensitive to the closed channel inclusion. This prediction given in
Ref.~\cite{YepezSaenz:2013} allows us to understand the intriguing numerical
results shown in Fig.~\ref{ScatteAmplit1}.

\begin{figure}
\includegraphics[scale=0.4]{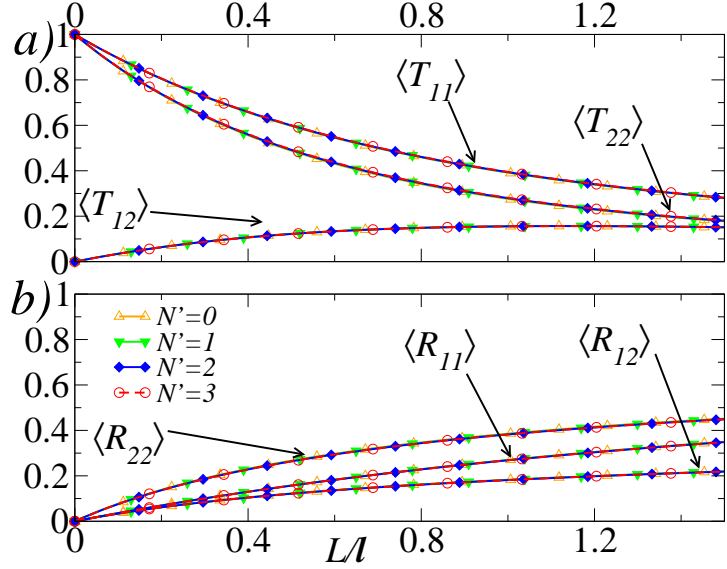}
\caption{Numerical results for $\left\langle T_{aa_{0}}\right\rangle$ and
$\left\langle R_{aa_{0}}\right\rangle$ vs $L/\ell$. Each simulation considers
$N=2$ propagating modes and different evanescent modes ($N^{\prime}=0,1,2,3$).}
\label{Ind_Taa0_Raa0}
\end{figure}

\begin{figure}
\includegraphics[scale=0.295]{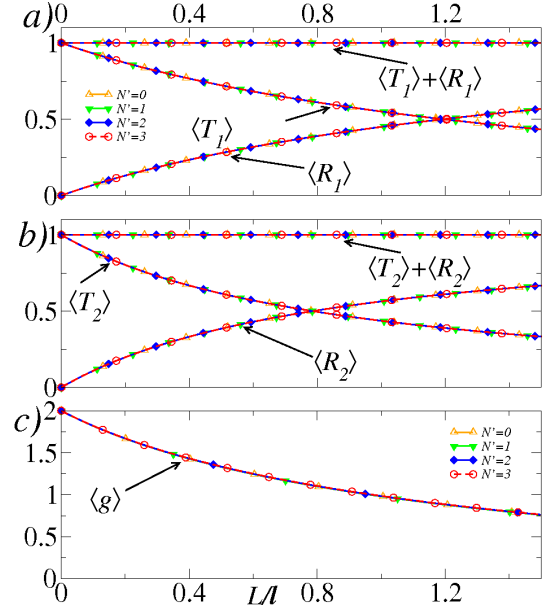}
\caption{Panels (a) and (b) shows the numerical expectation values of the total
transmittance $\left\langle T_{a_{0}}\right\rangle$ and reflectance 
$\left\langle R_{a_{0}}\right\rangle$; the flux conservation property
$\left\langle T_{a_{0}}\right\rangle+\left\langle R_{a_{0}}\right\rangle=1$ is
also shown. In (c) it is shown the numerical expectation value of the
dimensionless conductance $\left\langle g \right\rangle$. Each simulation
considers $N=2$ propagating modes and different evanescent modes
($N^{\prime}=0,1,2,3$).}
\label{FluxConseProp}
\end{figure}

In Fig.~\ref{Ind_Taa0_Raa0}, we present the complete set of numerical
expectation values of the transmittance $\left\langle T_{aa_{0}}\right\rangle$
and reflectance $\left\langle R_{aa_{0}}\right\rangle$. From this figure, we can
appreciate that the closed channels do not have any effect in the expectation
values $\left\langle T_{aa_{0}}\right\rangle$ and 
$\left\langle R_{aa_{0}}\right\rangle$. In a similar way,
Fig.~\ref{FluxConseProp} shows that the expectation values of the total
transmittance 
$\left\langle T_{a_{0}}\right\rangle=\sum_{a=1}^{N}\left\langle
T_{aa_{0}}\right\rangle$ and reflectance 
$\left\langle R_{a_{0}} \right\rangle=\sum_{a=1}^{N}\left\langle
R_{aa_{0}}\right\rangle$, as well as the dimensionless conductance 
$\left\langle g\right\rangle=\sum_{a,a_{0}=1}^{N}\left\langle
T_{aa_{0}}\right\rangle=\sum_{a_{0}=1}^{N}\left\langle T_{a_{0}}\right\rangle$,
are also insensitive to the closed channel contributions. In
Fig.~\ref{FluxConseProp}, we can also appreciate that the numerical expectation
values $\langle T_{a_{0}}\rangle$ and $\langle R_{a_{0}}\rangle$ satisfy the
flux conservation property, Eq.~\eqref{FluxConProp}, i.e., 
$\langle T_{a_{0}}\rangle+\langle R_{a_{0}}\rangle=1$; therefore, the numerical
results are consistent with the flux conservation property.

The numerical evidence shown in Figs.~\ref{Ind_Taa0_Raa0}
and~\ref{FluxConseProp}, confirms previous numerical results, where the closed
channels do not contribute in the statistics of transport
coefficients~\cite{PhysRevE.75.031113}. This evidence justifies the omission of
the closed channels in the description of the statistical properties of the
transport coefficients $T_{aa_{0}}$ and $R_{aa_{0}}$, which is a common
approximation in the theoretical study of wave transport through disordered
systems. The no role of the of the closed channels in the statistics of the
transport coefficients $T_{aa_{0}}$, $R_{aa_{0}}$ and $g$, is theoretically
explain in Ref.~\cite{YepezSaenz:2013}, where it is demonstrated that the
statistical properties of the transmittance and the reflectance only depends on
the scattering mean free path $\ell_{a_{0}}$, in which the closed channels play
no role.

\begin{figure}
\includegraphics[scale=0.295]{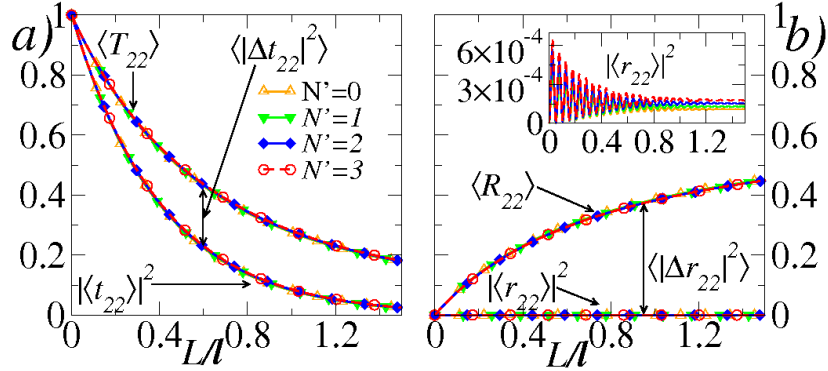}
\caption{Numerical results for (a) the transmittance 
$\left\langle T_{22}\right\rangle$ and (b) the reflectance 
$\left\langle R_{22}\right\rangle$, and their coherent intensities
$\vert\left\langle t_{22}\right\rangle\vert^{2}$, 
$\vert\left\langle r_{22}\right\rangle\vert^{2}$. Each simulation considers
$N=2$ propagating modes and different evanescent modes ($N^{\prime}=0,1,2,3$).}
\label{CoherentDiffuseIntensities}
\end{figure}

In contrast with the numerical results shown in Fig.~\ref{ScatteAmplit1} for the
coherent fields $\left\langle t_{aa_{0}}\right\rangle$ and 
$\left\langle r_{aa_{0}}\right\rangle$, the expectation values of the
transmittance $\left\langle T_{aa_{0}}\right\rangle$ and reflectance
$\left\langle R_{aa_{0}}\right\rangle$ are insensitive to the closed channel
influence, which means that both, coherent 
$\vert\langle t_{aa_{0}}\rangle\vert^{2}$, 
$\vert\langle r_{aa_{0}}\rangle\vert^{2}$ and diffuse 
$\langle\vert\Delta t_{aa_{0}}\vert^{2}\rangle$, 
$\langle\vert\Delta r_{aa_{0}}\vert ^{2}\rangle$ intensities, Eqs.
\eqref{Taa0Coehe_Diffuse_fiels}-\eqref{Raa0Coehe_Diffuse_fiels}, are insensitive
to the closed channel contributions. In order to analyze the closed channel
influence in both kinds of intensities, in Fig.~\ref{CoherentDiffuseIntensities}
we plotted the ensemble average of the transport coefficients 
$\langle T_{22}\rangle$, $\langle R_{22}\rangle$ and the corresponding coherent
intensities $\vert\langle t_{22}\rangle\vert^{2}$, 
$\vert\langle r_{22}\rangle\vert^{2}$. Fig.~\ref{CoherentDiffuseIntensities}(a)
shows that the coherent field intensity $\vert\langle t_{22}\rangle\vert^{2}$
becomes less important for large values of $L/\ell$, where the diffuse field
governs the behavior of $\langle T_{22}\rangle$. In contrast,
Fig.~\ref{CoherentDiffuseIntensities}(b) shows that the diffuse field 
$\langle\vert\Delta r_{22}\vert^{2}\rangle$ dominates the behavior of 
$\langle R_{22} \rangle$ for all $L/\ell$.

\section{Conclusions}
\label{Sec:Conclusions}

We analyzed numerically the influence of the closed channels in the statistical
scattering properties of disordered waveguides. Our numerical results show that
the statistical average of the transmittance 
$\left\langle T_{aa_{0}}\right\rangle$, the reflectance 
$\left\langle R_{aa_{0}}\right\rangle$, and the dimensionless conductance
$\left\langle g\right\rangle$ are insensitive to the closed channels inclusion,
which is in good agreement with previous numerical simulations. In contrast, the
expectation values of complex transmission and reflection coefficients, i.e.,
the coherent fields $\left\langle t_{aa_{0}}\right\rangle$, 
$\left\langle r_{aa_{0}}\right\rangle$ depend drastically on the closed channel
contributions. Since the numerical results show that the coherent intensities 
$\vert\langle t_{aa_{0}}\rangle\vert^{2}$, 
$\vert\langle r_{aa_{0}}\rangle\vert^{2}$ do not depend strongly on the closed
channels inclusion, then closed channels modify mainly the phases of the
coherent fields $\left\langle t_{aa_{0}}\right\rangle$, 
$\left\langle r_{aa_{0}}\right\rangle$.

The numerical results also exhibit that the transmittance 
$\left\langle T_{aa_{0}}\right\rangle$ and the reflectance 
$\left\langle R_{aa_{0}}\right\rangle$ are mainly dominated by their
corresponding diffuse $\langle\vert\Delta t_{aa_{0}}\vert^{2}\rangle$,
$\langle\vert\Delta r_{aa_{0}}\vert^{2}\rangle$ intensities: the coherent
intensity of the transmittance $\vert\langle t_{aa_{0}}\rangle\vert^{2}$,
decreases rapidly as the length $L$ of the disordered region increases, while
the coherent intensity of the reflectance 
$\vert\langle r_{aa_{0}}\rangle\vert^{2}$ is negligible for any value of $L$.

\begin{theacknowledgments}

The author thanks J. Feilhauer, L. Froufe-P\'erez, J.~J. S\'aenz and P.~A.
Mello for important discussions, and C. Lopez Nataren for technical support in
the numerical simulations. This work has been supported by postdoctoral grants
(No. 162768 and 187138) of the Mexican Consejo Nacional de Ciencia y
Tecnolog\'ia. 

\end{theacknowledgments}

\bibliographystyle{aipproc}
\bibliography{AIPConfProcBibYepez}

\end{document}